\documentclass[11pt]{article}
\usepackage{RR}
\RRNo{6529}

\usepackage{latexsym,amssymb}
\usepackage{graphicx,epsfig}
\usepackage{theorem,subfigure}
\usepackage{algorithm}
\newfloat{Algorithm}{hpt}{lop}

\newtheorem{theorem}{Theorem}

\newtheorem{proposition}{Proposition}

\def\Proof{\par\noindent{\bf Proof:}\indent}
\def\QED{\hfill$\Box$\par\vskip1em}

\RRdate{May 2008}
\RRtitle{D\'{e}tecteurs d'Univers pour la D\'{e}fense contre les Attaques Sybilles dans les R\'{e}seaux Ad Hoc Sans Fil}
\RRetitle{Universe Detectors for Sybil Defense in Ad Hoc Wireless Networks}
\titlehead{Universe Detectors for Sybil Defense}
\RRauthor{Adnan Vora$^{\dagger}$ \and Mikhail Nesterenko$^{\dagger}$ \and S\'{e}bastien Tixeuil$^\star$ \and Sylvie Dela\"{e}t$^\ddag$\\
\vspace{.5cm}
\small{
$^\dagger$ Dept. of Computer Science, Kent State University, USA\\
$^\star$ Universit\'{e} Pierre et Marie Curie-Paris 6, LIP6-CNRS, INRIA Grand Large, France\\
$^{\ddag}$ Universit\'{e} Paris Sud-XI, LRI-CNRS, France
}
}

\authorhead{A. Vora \emph{et al.}}

\RRresume{
Le probl\`{e}me de l'attaque Sybille dans les r\'{e}seaux \`{a} ports inconus comme les r\'{e}seaux sans fil n'est pas consid\'{e}r\'{e} comme soluble. Un n\oe ud sans fil n'est pas capable de diff\'{e}rencier par lui m\^{e}me un univers de n\oe uds r\'{e}els d'un univers de n\oe uds fictifs cr\'{e}\'{e} par un attaquant.

De mani\`{e}re similaires aux d\'{e}tecteurs de d\'{e}faillances, nous proposons d'utiliser des \emph{d\'{e}tecteurs d'univers} pour aider les n\oe uds \`{a} d\'{e}terminer quel univers est r\'{e}el. Dans cet article, nous \emph{(i)} d\'{e}finissons plusieurs variantes du probl\`{e}me de d\'{e}couverte de voisinage en pr\'{e}sence d'attaques Sybilles; \emph{(ii)} nous pr\'{e}sentons un ensemble de d\'{e}tecteurs d'univers correspondants; \emph{(iii)} nous prouvons la n\'{e}cessit\'{e} d'utiliser des contraintes topologiques suppl\'{e}mentaires pour que le probl\`{e}me devienne soluble: la densit\'{e} des n\oe uds et la port\'{e}e de communication; \emph{(iv)} nous pr\'{e}sentons $\cal SAND$, un algorithme distribu\'{e} qui r\'{e}soud les probl\`{e}mes propos\'{e}s \`{a} l'aide des d\'{e}tecteurs d'univers appropri\'{e}s, et montrons que les d\'{e}tecteurs d'univers sont les plus faibles possibles pour chaque probl\`{e}me. 
}

\RRabstract{
The Sybil attack in unknown port networks such as wireless is not
considered tractable. A wireless node is not capable of independently
differentiating the universe of real nodes from the universe of
arbitrary non-existent fictitious nodes created by the
attacker. Similar to failure detectors, we propose to use
\emph{universe detectors} to help nodes determine which universe is
real. In this paper, we (i) define several variants of the
neighborhood discovery problem under Sybil attack (ii) propose a set
of matching universe detectors (iii) demonstrate the necessity of
additional topological constraints for the problems to be solvable:
node density and communication range; (iv) present $\cal SAND$ --- an
algorithm that solves these problems with the help of appropriate
universe detectors, this solution demonstrates that the proposed
universe detectors are the weakest detectors possible for each
problem.
}

\RRmotcle{Attaque Sybille, r\'{e}seau sans fil, d\'{e}tecteur d'univers
}
\RRkeyword{Sybil attack, wireles network, unverse detector
}
\RRprojet{Grand Large}
\RRtheme{\THNum}
\URFuturs

\begin{document}
\makeRR

\section{Introduction} 

A Sybil attack, formulated by Douceur~\cite{D02}, is intriguing in its
simplicity. However, such an attack can incur substantial damage to
the computer system. In a Sybil attack, the adversary is able to
compromise the system by creating an arbitrary number of identities
that the system perceives as separate. If the attack is successful,
the adversary may either overwhelm the system resources, thus
channeling the attack into denial-of-service~\cite{WS02}, or create
more sophisticated problems, e.g. routing infrastructure
breakdown~\cite{KW03}.

Ad hoc wireless networks, such as a sensor networks, are a potential
Sybil attack target. The ad hoc nature of such networks may result in
scenarios where each node starts its operation without the knowledge
of even its immediate neighborhood let alone the complete network
topology. Yet, the broadcast nature of the wireless communication
prevents each node from recognizing whether the messages that it
receives are sent by the same or different senders. Thus, an attacker
may be free to either create an arbitrary number of fictitious
identities or impersonate already existing real nodes. The problem
straddles the security and fault tolerance domains as the attacker may
be either a malicious intruder or a node experiencing Byzantine
fault. A fault is Buzantine~\cite{LSP82} if the faulty node disregards
the program code and behaves arbitrarily. For convenience,
in this paper we assume that the attacker is a faulty node rather than
intruder.

\noindent
\textbf{Problem motivation.}  A standard way of establishing
trust between communicating parties is by employing
cryptography. There is a number of publications addressing the Sybil
attack in this manner
\cite{DHM03,pseudonyms,PPG05,TB06,YYYLA05,ZWRN05,ZSJ03}.  For example,
if each node has access to verified certificates and every sender
digitally signs its messages, then the receiver can unambiguously
determine the sender and discard superfluous identities created by the
faulty node by checking the digital signature of the message against
the certificates. However, there are several reasons for this approach
to be inappropriate. A cryptography-based solution pre-supposes a
key-based infrastructure which requires its maintenance and update and
thus limits its applicability.  Moreover, resource constrained
devices, such as sensor nodes in sensor networks, may not be able to
handle cryptographic operations altogether.

Another approach is intrusion detection based on
\emph{reputation}~\cite{BB03,CF05,JMB03}.  Due to the broadcast nature
of wireless communication, the messages from each node are observed by
its neighbors. A fault is detected if the node deviates from the
protocol. It is unclear how reputation-based schemes would fare if the
messages cannot be matched to the sender: the faulty node may
impersonate other nodes or create an arbitrary number of fictitious
nodes and set up its own alternative reputation verification network.

However, there are two unique features of wireless communication that
make defense against the Sybil attack possible. The wireless
communication is broadcast. Thus, the message transmission of a faulty
node is received by all nodes in its vicinity. In addition, the nodes
can estimate the \emph{received signal strength} (RSS) of the message
and make judgments of the location of the sender on its basis. Note
that the latter is not straightforward as the faulty node can change
its \emph{transmission signal strength} (TSS). In this paper we
investigate the approaches to Sybil defense using this property of
wireless communication.

\noindent 
\textbf{Related literature.} Newsome et al~\cite{NSSP04} as well
as Shi and Perrig~\cite{SP04} survey various defenses against the
Sybil attack. They stress the promise of the type of technique we
consider. Demirbas and Song~\cite{DS06} consider using the RSS for
Sybil defense.

A line of inquiry that is related to Sybil defense is secure location
identification~\cite{CH06,KZ03,LPC05,SSW03,VN06}. In this case, a set
of trusted nodes attempt to verify the location of a possibly
malicious or faulty node. However, the establishment of such trusted
network is not addressed. Hence, this approach may not be useful for
Sybil defense.

Dela{\"e}t et al~\cite{DMRT08c}, and Hwang et al~\cite{HHK07} consider
the problem where the faulty node operates synchronously with the
other nodes.  Dela{\"e}t et al~\cite{DMRT08c} provided examples of
positioning of faulty nodes and their strategies that lead to
neighborhood discovery compromise. Note that the synchrony assumption
places a bound on the number of distinct identities that the faulty
node can assume before the correct nodes begin to counter its
activities. Even though the faulty node may potentially create the
infinite number of fictitious identities, the correct nodes have to
deal with no more than several of them at a time. However, this
approach simplifies the problem as it limits the power of the faulty
node and the strength of the attack.

Nesterenko and Tixeuil~\cite{NT06} describe how, despite Byzantine
faults, every node can determine the complete topology of the network
despite once each node recognizes its immediate neighbors. Thus, to
defend against the Sybil attack it is sufficient to locally solve
Byzantine-robust neighborhood discovery.

Note that the problem is trivial when the ports are known. In this
case, the receiver may not know the identity of the transmitter of the
message but can match the same transmitter across messages. This
prohibits the faulty node from creating more than a single fictitious
identity or impersonating other real nodes and allows a simple
solution.

\noindent
\textbf{Our approach and contribution.} We consider the problem
of neighbor identification in the presence of Byzantine nodes. The
nodes are embedded in a geometric plane and know their location. They
do not have access to cryptographic operations. The nodes can exchange
arbitrary messages, but the only information about the message that
the receiver can reliably obtain is its RSS. We consider the
asynchronous model of execution. That is, the execution speed of any
pair of nodes in the network can differ arbitrarily. This enables the
faulty node to create an arbitrary number of fictitious identities or
impersonate the correct nodes in an arbitrary way. Moreover, in this
model, the only unique identities that the nodes have are their
coordinates. Hence, the objective of each node is to collect the
coordinates of its neighbors. We focus on local solutions to the
neighborhood discovery. That is, each node only processes messages
from the correct neighbors within a certain fixed distance. We do not
consider a denial-of-service attack or jamming attack~\cite{WS02},
where the faulty nodes just overwhelm resources of the system by
continuously transmitting arbitrary messages. We assume that the
network has sufficient bandwidth for message exchanges and the nodes
have sufficient memory and computing resources to process them.  To
the best of our knowledge, this is the most general model of Sybil
defense considered to-date.

In Section~\ref{SecModel} we provide details for our execution model
and formally state several variants of the neighborhood discovery
problem. Sections~\ref{SecNotSolvable}, \ref{SecDetect},
\ref{SecDensity}, and \ref{SecRange} outline the boundaries of the
achievable.  In Section~\ref{SecNotSolvable}, we formally prove that
this problem is not solvable without outside help. Intuitively, the
faulty node may create a \emph{universe} of an arbitrary number of
fictitious identities whose messages are internally consistent and the
correct node has no way of differentiating those from the universe of
correct nodes.  In Section~\ref{SecDetect}, we introduce
\emph{universe detectors} as a way to help nodes select the correct
universe. The idea is patterned after failure
detectors~\cite{CT96}. Just like failure detectors, universe detectors
are not implementable in asynchronous systems. However, they provide a
convenient abstraction that separates the concerns of algorithm design
and implementation of the necessary synchrony and other details that
enable the solution to Sybil defense. However, unlike failure
detectors, universe detectors alone are insufficient to allow a
solution to the neighborhood discovery problem. If the density of the
network is too sparse, the faulty nodes may introduce a fictitious
identity such that the detector is rendered unable to help the correct
nodes. In Section~\ref{SecDensity}, we prove the necessary condition
for the location of the correct nodes to allow a solution to the
neighborhood discovery problem. However, the faulty node may still be
able to compromise the operation of correct nodes. For that, a faulty
node may assume the identity of a correct node and discredit it by
sending incorrect messages to other nodes. In Section~\ref{SecRange}
we prove another necessary condition for the minimum transmission
range of correct nodes that eliminates this problem.

In Section~\ref{SecSAND} we present a Sybil-attack resilient
neighborhood discovery algorithm $\cal SAND$ that uses the universe
detectors to solve the neighborhood discovery problem provided that
the necessary conditions are met.  In their study of failure
detectors Chandra et al~\cite{CHT96} defined the weakest failure
detector as the necessary detector to solve the problem that they are
deployed to address. With the introduction of $\cal SAND$, we show
that the employed detectors are the weakest detectors necessary to
solve the neighborhood discovery problem. In Section~\ref{SecEnd}, we
conclude the paper by discussing the implementation details of the
algorithm and the attendant universe detectors.

\section{Computation Model  Description, Assumptions,
         Notation and Definitions}\label{SecModel}

A computer network consists of nodes embedded in a geometric
plane. Each node is aware of its own coordinates.  A (\emph{node})
\emph{layout} is a particular set of nodes and their locations on the
plane. Unless explicitly restricted, we assume that the node layout
can be arbitrary. Any specific point on the plane can be occupied by
at most one node. Thus, the node's coordinates on the plane uniquely
identify it. The nodes have no other identifiers.  For ease of
exposition, we use identifiers at the end of the alphabet such as $u$
or $v$ to refer to the particular locations or non-faulty nodes
occupying them. We use $f$ and $k$ respectively to refer to a faulty
node and a location where the faulty node may pretend to be located.
The distance between $u$ and $v$ is $|uv|$. The \emph{neighborhood
set} or just \emph{neighborhood} of a node $u$ is a set of nodes whose
distance to $u$ is less than a certain fixed distance $d_n$.

\noindent
\textbf{Program model.}  We assume the asynchronous model of
algorithm execution. That is, the difference between the execution
speed of any pair of nodes can be arbitrarily large.  Note that this
asynchrony assumption allows any node, including a faulty one, to send
an arbitrary number of messages before other nodes are able to
respond. The nodes run a distributed algorithm. The algorithm consists
of variables and actions. A (\emph{global}) \emph{state} of the
algorithm is an assignment of values to all its variables.  An action
is \emph{enabled} in a state if it can be executed at this state. A
\emph{computation} is a maximal fair sequence of algorithm states
starting from a certain prescribed initial state $s_0$ such that for
each state $s_i$, the next state $s_{i+1}$ is obtained by atomically
executing an action that is enabled in $s_i$. Maximality of a
computation means that the computation is either infinite or
terminates where none of the actions are enabled. In other words, a
computation cannot be a proper prefix of another computation. Fairness
means that if an action is enabled in all but finitely many states of
an infinite computation then this action is executed infinitely
often. That is, we assume \emph{weak fairness} of action
execution. During a single computation, node layout is fixed.

Nodes can be either correct or faulty (Byzantine). A faulty node does
not have to follow the steps of the algorithm and can behave
arbitrarily throughout the computation.

\noindent
\textbf{Node communication.}  Nodes communicate by broadcasting
messages. As the distance to the sender increases, the signal
fades. We assume the free space model~\cite{Rappaport02} of signal
propagation. The antennas are omnidirectional. The received signal
strength (RSS) changes as follows:
\begin{equation}\label{eqnSignalPropagation}
R=cT/r^2
\end{equation}
where $R$ is the RSS, $c$ is a constant, $T$ is the transmitted (or
sent) signal strength (TSS), and $r$ is the distance from the sender
to the receiver. We assume that $r$ cannot be arbitrarily small. Thus,
$R$ is always finite. There is a minimum signal strength $R_{min}$ at
which the message can still be received. There is no message
loss. That is, if a message is sent with TSS --- $T'$, then every node
within distance $r'=\sqrt{cT'/R_{min}}$ of the sender receives the
message. We do not consider interference, hidden-terminal effect or
other causes of message loss. We assume that every correct node always
broadcasts with a certain fixed strength $T_r$. A \emph{range} $r_t$
is defined as $\sqrt{cT_r/R_{min}}$. The relation between range $r_t$
and neighborhood distance $d_n$ is, in general, arbitrary. A faulty
node may select arbitrary TSS. If a node receives a message (i.e. if
the RSS is greater than $R_{min}$), then the node can accurately
measure the RSS.

To simplify the exposition we assume that the nodes transmit three
types of messages: (i) $u$ transmits \emph{announce}, this message has
only the information about $u$ and carries $u$'s coordinates; the
purpose of an announcement is for $u$ to advertise its presence to its
neighbors; (ii) $u$ transmits \emph{confirm} of another node $v$'s
transmission; (iii) $u$ transmits \emph{conflict} with another node
$v$'s transmission if its observations do not match the location or
the contents of $v$'s message. The original message is attached in
\emph{confirm} and \emph{conflict}. Every message contains the
coordinates of the sender.

\noindent
\textbf{Fictitious nodes and conflicts.} Since the only way to
unambiguously differentiate the nodes is by their location, the
objective of every node is to determine the coordinates of its
neighbors. Faulty nodes may try to disrupt this process by making the
correct node assume that it has a non-existent neighbor. Such a
non-existent neighbor is \emph{fictitious}.  A node that indeed exists
in the layout is \emph{real}. Note that a real node can still be
either correct or faulty. Faulty nodes may try to tune their TSS and
otherwise transmit messages such that it appears to the correct nodes
that the message comes from a fictitious node. Moreover, the faulty
nodes may try to make their transmissions appear to have come from
correct nodes.

As a node receives messages, due to the actions of a faulty node, the
collected information may be contradictory. A \emph{conflict} consists
of a message of any type purportedly coming from node $k$, yet the
received signal strength at node $u$ does not match $|uk|$ provided
that the signal were broadcast from $k$ with the TSS of $T_r$.  A
conflict is \emph{explicit} if $u$ receives this conflicting message.
Note that the RSS may be so low that $u$ is unable to receive the
message altogether, even though the RSS at $u$ should be greater than
$R_{min}$ in case the message indeed come from $k$ and be broadcast at
$T_r$. In this case the conflict is \emph{implicit}.  To discover the
implicit conflict $u$ has to consult other nodes that received the
conflicting message. If $u$ detects a conflict it sends a conflict
message.

A \emph{universe} is a subset of neighbors that do not conflict. That
is, a universe at node $u$ contains nodes $v$ and $w$ whose
announcements $u$ received such that $u$ did not receive a conflict
from $v$ about $w$ or from $w$ about $v$.  Note that due to conflicts
the information collected by a single node may result in several
different universes. A universe is \emph{real} if all nodes in it are
real.  A universe is \emph{complete} for a node $u$ if it contains all
of $u$'s correct neighbors. Note that even though a faulty node is
real, it can evade being added to universes by not sending any
messages. Hence, a complete universe is not required to contain all
the real nodes, just correct ones.

\noindent
\textbf{Program locality.}  To preserve the locality of a
solution to the neighborhood discovery problem, we introduce the
following requirement. Each node ignores information from the nodes
outside the range $r_t$ and about the nodes outside the neighborhood
distance $d_n$.

\noindent
\textbf{Problem statement.} We define several variants of the
problem.  The \emph{strong neighborhood discovery problem} $\cal SNDP$
requires each correct node $u$ to output its neighborhood set
according to the following properties:

\begin{description}
\item[safety] --- if the neighborhood set of $u$ is output, the set
  contains only all correct nodes and no fictitious nodes of $u$'s
  neighborhood;

\item[liveness] --- every computation has a suffix in whose every
state $u$ outputs a neighborhood set that contains all correct
neighbors of $u$. In other words, $u$ eventually outputs its complete
neighborhood set.
\end{description}

This problem definition may be too strict. Some correct nodes may be
slow in announcing their presence. However, the safety property of
$\cal SNDP$ requires each node to wait for its slow neighbors before
outputting the neighborhood set.  Hence, we define the \emph{weak
neighborhood discovery problem} $\cal WNDP$. This problem relaxes the
safety property to allow the output neighborhood set to contain a
subset of correct neighbors of $u$. Note that the presence of the
fictitious nodes in the output is still prohibited. Also note that the
liveness property requires that the neighborhood set of $u$ in $\cal
WNDP$ eventually contains all correct neighbors.  Further relaxation
of the safety property yields the \emph{eventual neighborhood
discovery problem} $\diamond \cal NDP$. It requires that the safety of
$\cal SNDP$ be satisfied only in the suffix of a computation. That is
$\diamond \cal NDP$ allows the correct nodes to output incorrect
information arbitrarily long before providing correct output. Observe
that any solution to $\cal SNDP$ is also a solution to $\cal WNDP$,
and any solution to $\cal WNDP$ is also a solution to $\diamond \cal
NDP$.

\section{Impossibility of Standalone Solution to Neighborhood Discovery} 
\label{SecNotSolvable}

In this section we demonstrate that in the asynchronous system any
correct node is incapable of discovering its neighborhood if a
faulty node is present. The intuition for this result is as
follows. Since a faulty node is not restricted in the number of
messages that it generates, it can send an arbitrary number of
announcements introducing fictitious nodes. The faulty node can then
imitate arbitrary message traffic between these non-existent nodes.
On its own, a correct node is not able to differentiate these
fictitious nodes from the real ones.

\begin{theorem}\label{trmNotSolvable}
In an asynchronous system, none of the three variants of the
neighborhood discovery problem are deterministically solvable in the
presence of a single Byzantine fault.
\end{theorem}

\Proof We provide the proof for the eventual neighborhood discovery
problem. Since this problem is the weakest of the three that we
defined, the impossibility of its solution implies similar
impossibility for the other two.

Assume the opposite. Let $\cal A$ be a deterministic algorithm that
solves $\diamond\cal NDP$ in the presence of a faulty node.  Let us
consider an arbitrary layout $L_1$ that contains a faulty node
$f$. Let us consider another layout $L_2$ containing $f$ such that the
neighborhood $U_1$ in layout $L_1$ of at least one correct node $u$
differs from its neighborhood $U_2$ in $L_2$ and this difference
includes at least one correct node. Without loss of generality we can
assume that there exists a correct node $v$ such that $v\in U_1$ and
$v \not \in U_2$.

We construct two computations of $\cal A$: $\sigma_1$ on layout $L_1$
and $\sigma_2$ on layout $L_2$. The construction proceeds by
iteratively enlarging the prefixes of the two computations. In each
iteration, we consider the last state of the prefix of $\sigma_1$
constructed so far and find the action that was enabled for the
longest number of consequent steps. If there are several such actions,
we choose one arbitrarily. We attach the execution of this action to
the prefix of $\sigma_1$. If this action is a message transmission of
a node $w$ such that $w \in U_1$, we also attach the following action
execution to the prefix of $\sigma_2$: node $f$ sends exactly the same
message as $w$ in $\sigma_1$ with the TSS selected as $T=T_r
|uf|^2/|uw|^2$. Observe that $u$ receives the same message and with
the same RSS in this step of $\sigma_2$ as in the step added to
$\sigma_1$.  If the new action attached to $\sigma_1$ prefix is not by
a node in $U_1$, or it is not a message transmission, no action is
attached to the prefix of $\sigma_1$. We perform similar operations to
the prefix of $\sigma_2$. 

We continue this iterative process until maximal computations
$\sigma_1$ and $\sigma_2$ are obtained. Observe that by construction,
both computations are weakly fair computations of $\cal A$. Moreover,
in both cases $u$ receives exactly the same messages with exactly the
same RSS.

By assumption, $\cal A$ is a solution to $\diamond\cal NDP$. According
to the liveness property of the problem, $\sigma_1$ has a suffix where
$u$ outputs its neighborhood in every state and, due to the liveness
property, $\sigma_1$ contains a suffix where $u$'s neighborhood set
contains all correct nodes. In layout $L_1$ of $\sigma_1$, $v$ is
$u$'s correct neighbor. Hence, $v$ has to be included in this
set. That is, there is a suffix of $\sigma_1$ where $u$ outputs a
neighborhood set that contains $v$.  However, $u$ receives the same
messages in $\sigma_2$.  Since $\cal A$ is deterministic, $u$ has to
output exactly the same set in $\sigma_2$ as well. That is, $\sigma_2$
contains a suffix where the neighborhood set also contains $v$.
However, $v$ is fictitious in layout $L_2$ of $\sigma_2$. According to
the safety property of $\diamond\cal NDP$, every computation should
contain a suffix where the neighborhood set of $u$ excludes fictitious
nodes. That is, $\sigma_2$ of $\cal A$ violates the safety of
$\diamond\cal NDP$.  Hence, our assumption that $\cal A$ is a solution
to the weak neighborhood discovery problem is incorrect. The theorem
follows.  \QED

\section{Abstract Universe Detectors}\label{SecDetect}

\textbf{Definitions.}  The negative result of
Theorem~\ref{trmNotSolvable} hinges on the ability of a faulty node to
introduce an arbitrary number of fictitious nodes. A correct node
cannot distinguish them from its real neighbors. Still, a correct node
may be able to detect conflicts between nodes and separate them into
universes. However, it needs help deciding which universe is
real. This leads us to introduce the concept of a universe detector
that enables the solution to the neighborhood discovery problem in the
asynchronous computation model. A \emph{universe detector} indicates
to each correct node which universe is real.  It takes the universes
collected by the node as input and outputs which universe contains
only real nodes. That is, a universe detector \emph{points} to the
real universe. Note that the algorithm still has to collect the
neighborhood information and separate them into universes such that at
least one of them is real. If the algorithm does not provide a real
universe, the detector does not help.

Depending on the quality of the output, we define the following
detector classes. For each node $u$, a \emph{strongly perfect universe
detector} $\cal SPU$ has the following properties:
\begin{description}
\item[completeness] --- if a computation contains a suffix where in
     every state, $u$ outputs a real and complete universe, then this
     computation also contains a suffix where $\cal SPU$ at $u$ points
     to it;
\item[accuracy] --- if $\cal SPU$ points to a universe, this
     universe is real and complete.
\end{description}

The strongly perfect universe detector may be too restrictive or too
difficult to implement.  Unlike $\cal SPU$, a \emph{weakly perfect
universe detector} $\cal WPU$ may point to a real universe even if it
is not complete. That is, the definition of accuracy is relaxed to
allow the detector to point to a real universe that is not
complete. Note that $\cal WPU$ still satisfies the completeness
property and has to eventually point to the real universe if it is
available. A further relaxation of completeness and accuracy yields an
\emph{eventually perfect universe detector} $\diamond {\cal PU}$ which
satisfies both properties in a suffix of every computation. Observe
that the relationship between these detector classes is as follows:
${\cal SPU} \subset {\cal WPU} \subset \diamond {\cal PU}$

Observe that these universe detectors enable a trivial solution to the
neighborhood discovery problems: each node composes a universe for
every possible combination of the nodes that claim to be in its
neighborhood.  Naturally, as the node receives announcements from all
its correct neighbors, one of these universes is bound to be real and
complete. Hence, the detector can point to it. However, such an
approach essentially shifts the burden of separating fictitious and
real nodes to the detector while we are interested in minimizing the
detector's involvement. This leads us to introduce an additional
property of the algorithms that we consider. An algorithm that solves
the neighborhood discovery problem that uses detectors is
\emph{conflict-aware} if for each universe $U$ of node $u$, if nodes
$v$ and $w$ do not have a conflict and $v$ belongs to $U$ then $w$
also belongs to $U$. That is, the algorithm does not gratuitously
separate non-conflicting neighbors into different universes. In what
follows we focus on conflict-aware solutions.

\section{Necessary Node Density}\label{SecDensity}

Theorem~\ref{trmNotSolvable} demonstrates that to solve the
neighborhood discovery problem, any algorithm requires outside help
from a construct like a universe detector. However, the availability
of a universe detector may not be sufficient. Faulty nodes may take
advantage of a layout to announce a fictitious node without generating
conflicts. Then, a correct node running a conflict aware algorithm
never removes this fictitious node from the real universe. A universe
detector then cannot point to such a universe.

To illustrate the idea we start with a sequence of fictitious node
placement examples.

\subsection{Fictitious Nodes Placement Examples}

For this discussion we consider the neighborhood of a certain correct
node $u$ and a faulty node $f$ that tries to compromise $u$'s
neighborhood discovery. We denote $x$, $y$, $z$ --- the correct nodes
in the neighborhood of $u$ that that are respectively first, second
and third nearest to $f$. Note that to affect $u$, the faulty node $f$
does not itself have to be the neighbor of $u$. Our analysis proceeds
according to the number of correct receivers of messages sent by $f$.

\begin{figure}
\centering
\epsfig{figure=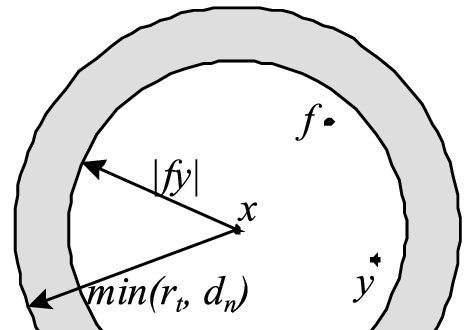,width=.7\textwidth,clip=}
\caption{Deception field with a single correct node
$x$.}\label{figOneReceiver}
\end{figure}

\noindent
\textbf{Single correct receiver.}  Refer to
Figure~\ref{figOneReceiver} for illustration. Note that due to the
broadcast nature of radio signal propagation, if any correct node
receives a message sent by $f$, $x$ also receives this message because
it is closest to $f$. Therefore, the single correct receiver may only
be $x$. Note, that for $y$ to not receive the signal from $f$, the
transmission signal strength should be sufficiently low. Recall that a
correct node always broadcasts with pre-defined signal strength
$T_r$. Thus, to deceive $x$, $f$ has to select the location of $k$ and
the TSS such that: (i) the RSS at $x$ is the same as if $k$
transmitted with $T_r$ and (ii) the RSS at $y$ is below $R_{min}$. For
the received signal strength at $y$ to be less than $R_{min}$, $k$
cannot be closer to $x$ than $|fy|$. On the other hand, the location
of $k$ cannot be outside the range $r_t$ (or else $x$ generates
conflict) or outside the neighborhood distance $d_n$ (or else $x$
ignores it). Thus, the possible location of $k$ is a ring around $x$
with the inner radius $|fy|$ and the outer radius --- $min(r_t,d_n)$.

\begin{figure}
\centering
\epsfig{figure=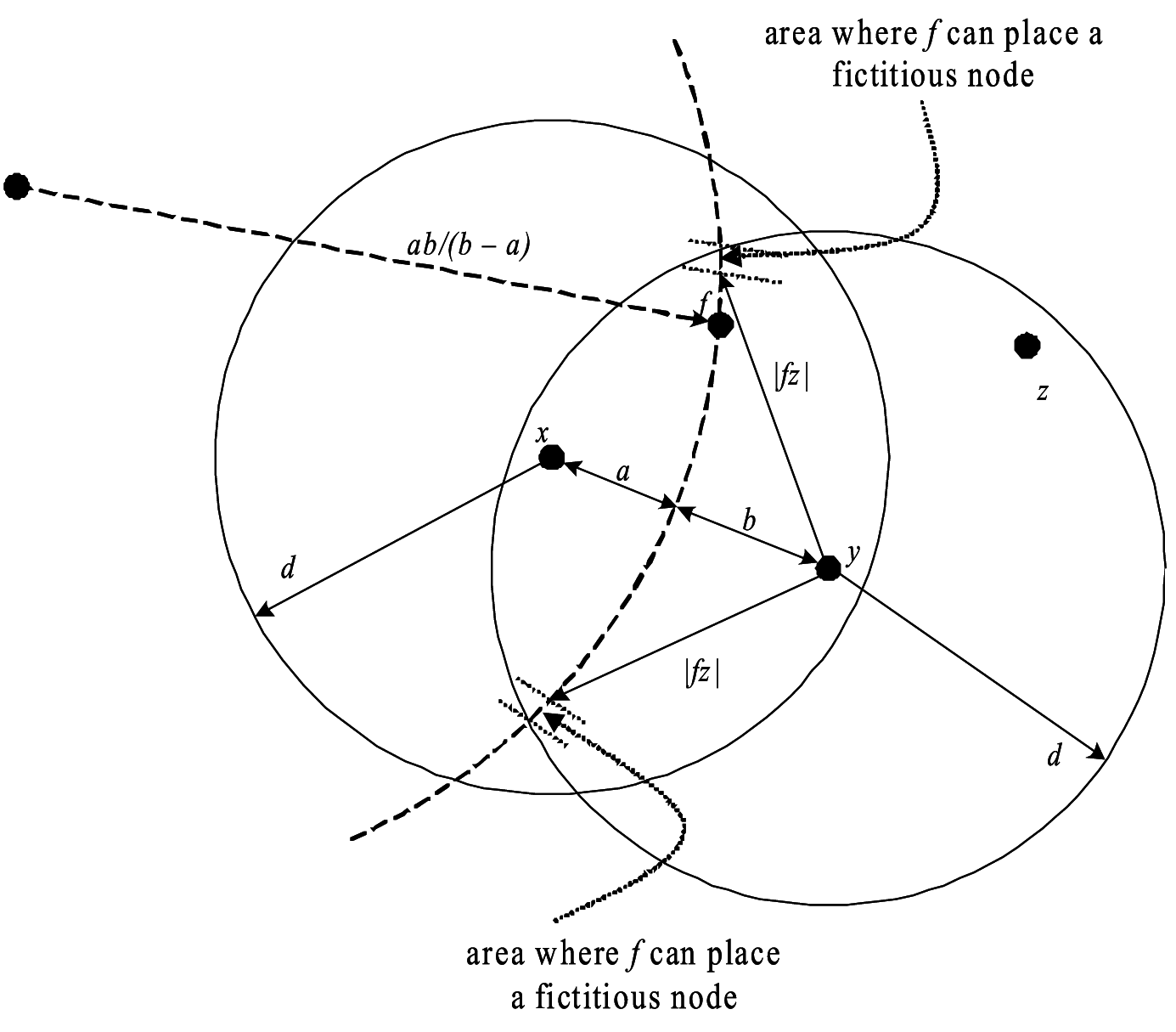,width=.7\textwidth,clip=}
\caption{Deception field with a two-node retinue. Correct nodes $x$
and $y$ receive transmissions of faulty node $f$, while $z$ does
not.}\label{figTwoReceivers}
\end{figure}

\noindent
\textbf{Two correct receivers.} If there are exactly two correct
receivers, they are the two nodes $x$ and $y$ nearest to $f$. Assume
that $f$ makes two transmissions at signal strengths $T_1$ and
$T_2$. For these transmissions, the RSS at $x$ and $y$ are $R_{x1}$,
$R_{y1}$ and $R_{x2}$, $R_{y2}$ respectively. From the signal
attenuation in Formula~\ref{eqnSignalPropagation} we obtain:
\[
 \frac{|fy|}{|fx|}=\sqrt{\frac{R_{x1}}{R_{y1}}}=
\sqrt{\frac{R_{x2}}{R_{y2}}}
\]
That is, regardless of transmission power, the ratio of received
signal strengths at $x$ and $y$ does not change. Hence, $f$ may select
the location of the fictitious node $k$ such that it preserves this
ratio. Such locations form an arc of a circle. Refer to
Figure~\ref{figTwoReceivers}. The center of the circle lies on the
line whose segment is $(xy)$. The radius of the circle is $ba/(b-a)$
where $b$ and $a$ are the portions of $(xy)$ such that
$b/a=|fy|/|fx|$. This circle is the \emph{deception circle}. Note that
$f$ may not be able to use all of the deception circle for fictitious
node placement: to get both $x$ and $y$ to receive the signal without
generating conflicts the points on the arc have to lie within
$min(r_t,d_n)$ of both $x$ and $y$.  Moreover, similar to the
single-receiver case, the portion of the arc that is closer to $y$
than $|fz|$ cannot be used without $z$ also receiving the message.
\begin{figure}
\centering
\epsfig{figure=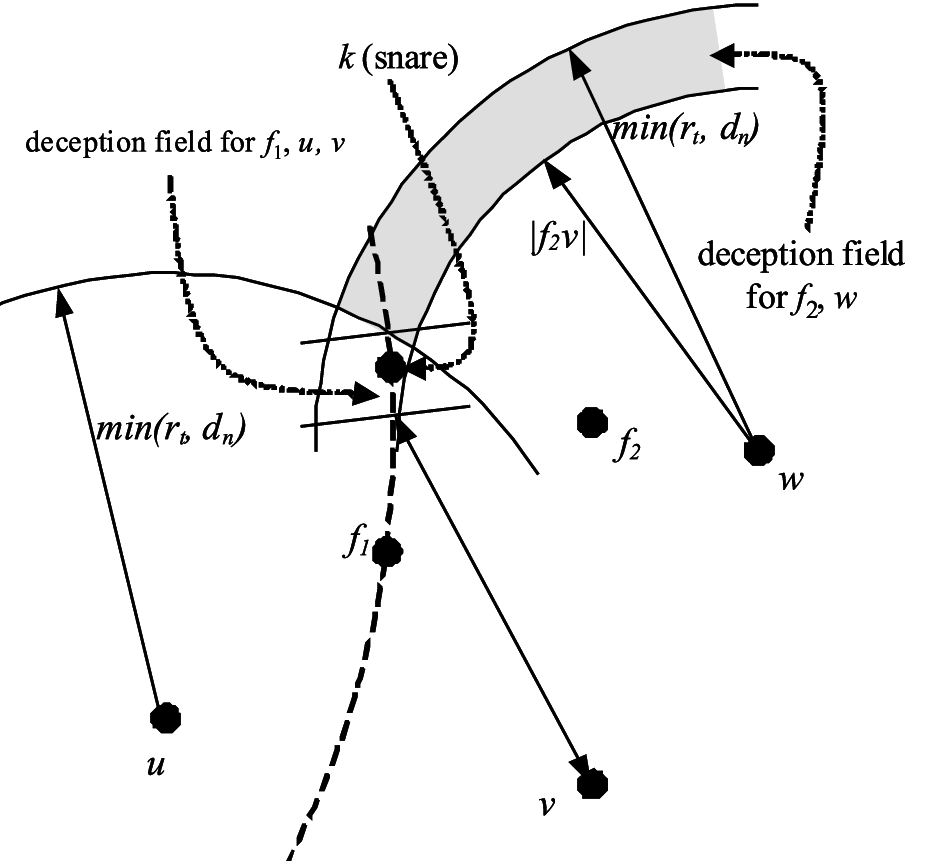,width=.7\textwidth,clip=}
\caption{The location of a snare in case of multiple faulty nodes. The
retinue of $f_1$ is $x$ and $y$. The retinue of $f_2$ is $z$. The
intersection of deception fields produces area where a snare can be
placed.}\label{figManyReceivers}
\end{figure}

\noindent
\textbf{More than two correct receivers.} Note that if there are
more than two correct receivers, they can be considered pairwise. Each
pair of correct receivers forms its own deception circle. Note that
$k$ can only be placed at the intersection of all these circles. Note,
however, that the circles intersect in the same place only if the
recipients are co-linear.

\noindent
\textbf{Snare.} A faulty node may affect the correct nodes around
it.  A set $E_f$ of correct nodes is the \emph{retinue} of a faulty
node $f$ if the following holds: if a correct node $u$ belongs to
$E_f$, then every correct node $v$ such that $|vf| \leq |uf|$, also
belongs to $E_f$. The faulty node is the \emph{leader} of the
retinue. For example, assume there are two faulty nodes $f_1$ and
$f_2$ and three correct nodes $u$, $v$ and $w$ such that $|f_1u| <
|f_1v| < |f_1w|$ and $|f_2w| < |f_2v| < |f_2u|$. Refer to
Figure~\ref{figManyReceivers} for illustration. All three correct
nodes can be either in the retinue $E_{f1}$ of $f_1$ or $E_{f2}$ of
$f_2$. However, if $v$ belongs of $E_{f1}$, so does $u$, and if $u$
belongs to $E_{f2}$, so do $v$ and $w$.

A \emph{deception field} for a retinue of a faulty node $f$ is the
area such that for each point $k$ of the field there exists a TSS that
the leader of the retinue can use to transmit a message. The message
so transmitted generates the RSS at each member of the retinue as if
the message was sent from $k$ with transmission strength
$T_r$. Intuitively, a deception field is the area where $f$ can place
fictitious nodes without generating conflicts at its retinue members.

A point $k$ in a neighborhood of a correct node $u$ is a
(\emph{simple}) \emph{snare} for $u$ if there exists a set of faulty
nodes and a retinue assignment for them such that: $u$ is in one of
the retinues and the intersection of the deception fields of the
retinues includes $k$. Note that the nodes in range of $k$ are either
in the retinues or not. Intuitively, a snare is a point where faulty
nodes can jointly place a fictitious node without generating explicit
conflicts at any of the correct neighbors of $u$.  Refer to
Figure~\ref{figManyReceivers} for illustration. Note that some of the
nodes may have implicit conflicts with $k$. That is, they are within
range $r_t$ of $k$ and $u$ but not in one of the retinues. That is,
they should receive a message from a node at $k$ but they do not. Note
that a snare transmission from faulty nodes may still generate
conflicts outside the range of $u$.  However, due to the locality
assumption, $u$ ignores this conflict.

A point $k$ is a \emph{perfect snare} for $u$ if it is a snare and all
nodes within the transmission range of $u$ and $k$ are in the retinues
of the faulty nodes participating in the snare.  That is, if faulty
nodes broadcast in a perfect snare, neither explicit nor implicit
conflicts are generated at the neighbors of $u$.

\begin{figure}
\centering
\epsfig{figure=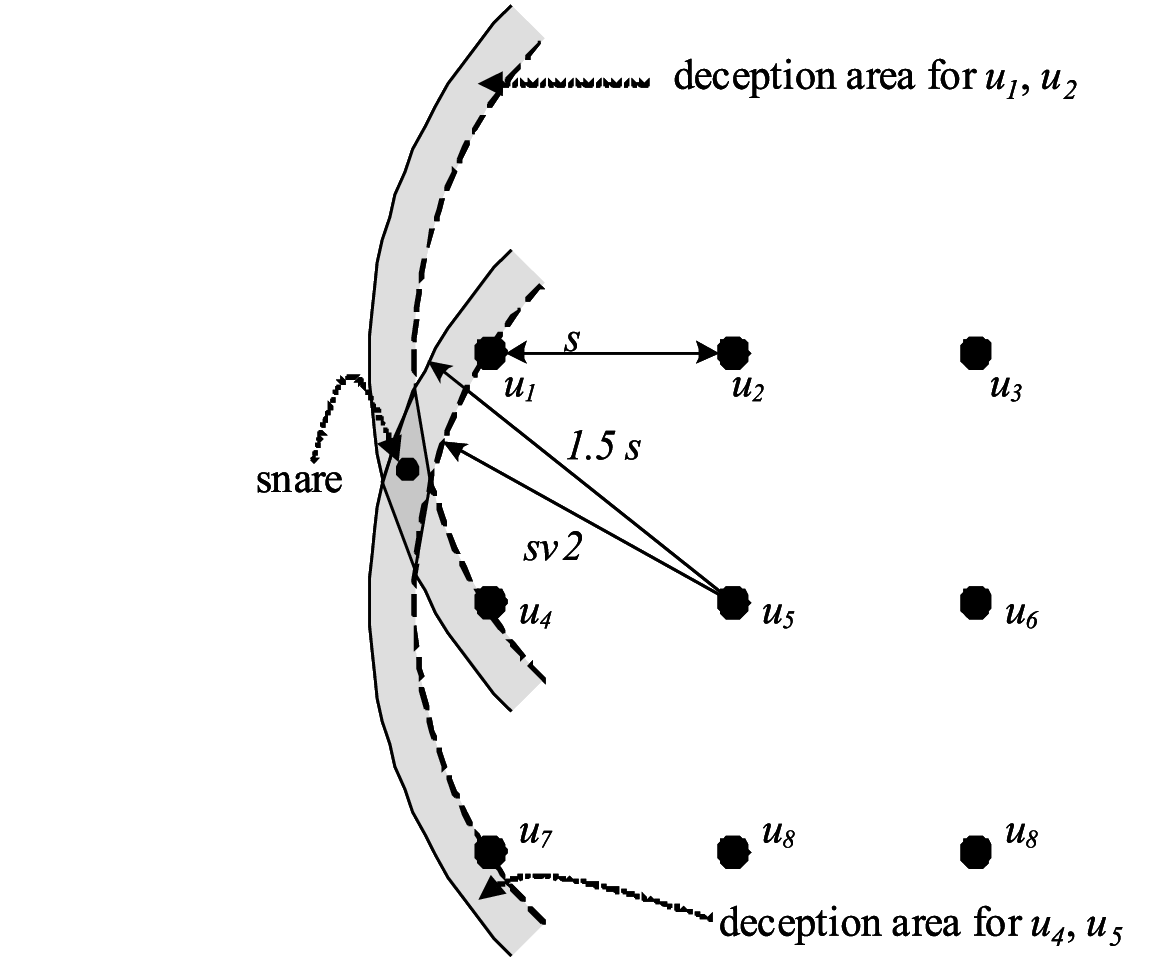,width=.7\textwidth,clip=}
\caption{The possibility of a snare in a grid layout with $d_n=1.5s$.}
\label{figGrid}
\end{figure}

\noindent
\textbf{Evaluating fault tolerance of a layout.} To illustrate
the concept of a snare, let us discuss a square grid layout (refer to
Figure~\ref{figGrid}). Let $s$ be the distance between the nodes in
the grid. Note that for a node to have any neighbors, the neighborhood
distance $d_n$ has to be no less than $s$. Let $ s \leq d_n < s
\sqrt{2}$. For simplicity, let $r_t = d_n$. Then each node has exactly
four neighbors. Note that the failure of a single node creates a
wedge-shaped deception field around the faulty node. Thus, with this
distance the layout is not fault tolerant.

Let us consider the case where $s\sqrt{2} \leq d_n < 2s$ and again
$r_t=d_n$. In this case, the neighborhood can withstand a failure of
exactly one node. Indeed, assume that a single node failed. Note that
we have to consider the collusion of this faulty node with arbitrary
faulty nodes outside the neighborhood. Let us focus on the
neighborhood of node $u_5$. For $u_5$ to consider a fictitious node,
the transmission of at least one faulty node has to reach
$u_5$. However, if a signal from a faulty node, either inside or
outside the neighborhood of $u_5$, reaches $u_5$, then this signal is
received by at least two more correct neighbors of $u_5$.  Moreover,
the three correct nodes that receive this signal are
non-collinear. This means that their pairwise deception circles
intersect only in the sender itself. Thus, the neighborhood of $u_5$
does not contain a snare.

Let us determine if this grid layout can withstand simultaneous
failure of two nodes in the same neighborhood.  Let $d_n=r_t$ be
$1.5s$. Suppose nodes $u_1$ and $u_4$ fail.  The deception field of
$u_1$ with $u_2$ in its retinue is a disk with outer circle radius
$1.5s$ and inner --- $s\sqrt{2}$. That is, the outer disk is the range
for correct nodes $r_t=d_n$ and the inner is the distance to the next
nearest node --- $u_5$. A similar disk is a deception field of $u_4$
for $u_5$. The intersection of the two disks forms the area where a
perfect snare for $u_5$ may be located. To use the snare, $u_1$ sends
messages to $u_2$, and $u_4$ to $u_5$ with the appropriate TSS
pretending that the messages come from a fictitious node located in
the snare. Thus, the grid layout with such $d_n$ cannot withstand a
two-node failure.

\subsection{Necessary Node Density Condition}

Having described the required instruments, we now demonstrate that the
availability of the universe detectors alone is not sufficient to
enable a solution to any of the neighborhood discovery problems if the
node layout is too sparse. That is, if the nodes are not properly
positioned on the plane.

To simplify the proof we consider solutions that are
\emph{well-formed}. An algorithm is well-formed if (i) the action that
transmits \emph{announcement} is always enabled until executed; (ii)
the receipt of a message may enable either \emph{confirm} or
\emph{conflict}, this action stays enabled until executed.

\begin{theorem} \label{trmNoSnare}
There is no conflict-aware well-formed deterministic solution to any of
the neighborhood discovery problems despite the availability of the
universe detectors if one of the considered layouts contains a perfect
snare.
\end{theorem}

\Proof In the proof, we focus again on the weakest of the problems:
the eventual neighborhood discovery.  Assume the opposite: there is a
conflict-aware well-formed algorithm $\cal A$ that uses a detector and
solves the problem even though in one of the layouts $L_1$, the
neighborhood of a correct node $u$ contains a perfect snare $k$.

Consider a layout $L_2$ that is identical to $L_1$ except that there
is a correct node at location $k$ in $L_2$. We construct a computation
$\sigma_2$ of $\cal A$ on $L_2$ as follows. Faulty nodes do not send
any messages in $\sigma_2$.  We arrange the neighbors of $u$,
including $u$ itself, into an arbitrary sequence $Q$. We then build
the prefix of $\sigma_2$ by iterating over this sequence. Since $\cal
A$ is well-formed, each node in the sequence has \emph{announcement}
enabled. We add the action execution that transmits
\emph{announcement} to $\sigma_2$ in the order of nodes in $Q$.  Since
$\cal A$ is well-formed, these transmissions may enable \emph{confirm}
actions at the neighbors of $u$. Note that since $v$ is correct,
\emph{conflict} actions are not enabled by these transmissions. We now
iterate over the nodes in $Q$. For each node $v$ we add the execution
of these \emph{confirm} actions at $v$ to $\sigma_2$ in arbitrary
fixed order, for example in the order that the original senders the
appear in $Q$.  We proceed in this manner until the sequence $Q$ is
exhausted. Note that these transmissions may potentially generate
another round of \emph{confirm} messages at the nodes in $Q$. We
continue iterating over $Q$ until no more messages are generated.  We
then complete $\sigma_2$ by executing the actions of nodes in an
arbitrary fair manner. Note that the remaining messages deal with the
nodes outside $u$'s neighborhood. Therefore, $u$ ignores them.

Now, the liveness property of all the detectors states that a detector
points to a universe if it is output for a suffix of the
computation. Since $\cal A$ is a solution of $\diamond\cal NDP$ and
$\sigma_2$ is a computation of $\cal A$, $\sigma_2$ has to contain a
suffix where $u$ outputs a real universe in every state. Since $k$ is
a correct neighbor of $u$, $k$ is included in the real universe.

Recall that in layout $L_1$, point $k$ is a perfect snare. This means
that there is an arrangement of retinues and the TSS for the faulty
nodes, such that when the faulty nodes transmit, each node in the
neighborhood of $u$ in the distance $d$ from $k$ receives a message
with the same RSS as if a node at $k$ broadcast with $T_d$. Moreover,
none of the nodes in the neighborhood of $u$ detect conflicts.

We construct a computation $\sigma_1$ of $\cal A$ on layout $L_1$ as
follows. We iterate over the same sequence $Q$ as in $\sigma_2$. Note
that $k$ is also present in the sequence even though it is fictitious
in $\sigma_1$. To build the prefix of $\sigma_1$ we execute similar
actions as for $\sigma_2$.  The only difference is that when node $k$
broadcasts in $\sigma_2$, in $\sigma_1$ we have the faulty nodes that
constitute the snare broadcast at the appropriate TSS. Note that in
the computation thus formed, the correct neighbors of $u$ receive
messages at the same RSS and with the same content from the faulty
nodes as in $\sigma_2$ from $k$. Thus, these transmissions do not
generate conflicts. Observe that this means that node $u$ receives the
same messages with the same RSS, and in the same sequence in
$\sigma_1$ and $\sigma_2$. Since $\cal A$ is deterministic, $u$ has to
output the same universes in $\sigma_1$ and $\sigma_2$. Note also,
that this means that $u$ does not record conflicts. Since $\cal A$ is
conflict aware, all $u$'s universes of $\cal A$ include $k$ together
with the correct neighbors.

However, $k$ is a fictitious node in $L_1$. This means that $\sigma_1$
contains a suffix where $u$ does not output a real universe. According
to the safety property of the detectors, none of them provides output
in a suffix of $\sigma_1$. Which means that $\cal A$ does not output a
neighborhood set in a suffix of $\sigma_1$. This violates the liveness
property of a solution to $\diamond\cal NDP$. Therefore, our
assumption that $\cal A$ is a solution to $\diamond\cal NDP$ is
incorrect. The theorem follows.\QED

\section{Necessary Transmission Range}\label{SecRange}

In this section we provide another required condition for the
existence of a solution to the neighborhood discovery
problem. Essentially, if the nodes in the same neighborhood are out of
range, the faulty node may introduce a conflict between them. This
forces the algorithm to mistakenly split the correct nodes into
separate universes and renders the failure detector powerless.

\begin{theorem}\label{trmRange}
There is no conflict-aware deterministic solution for any of the
neighborhood discovery problems despite the availability of universe
detectors and lack of snares if the node transmission range $r_t$ is
less than double the neighborhood distance $d_n$.
\end{theorem}

\begin{figure}
\centering
\epsfig{figure=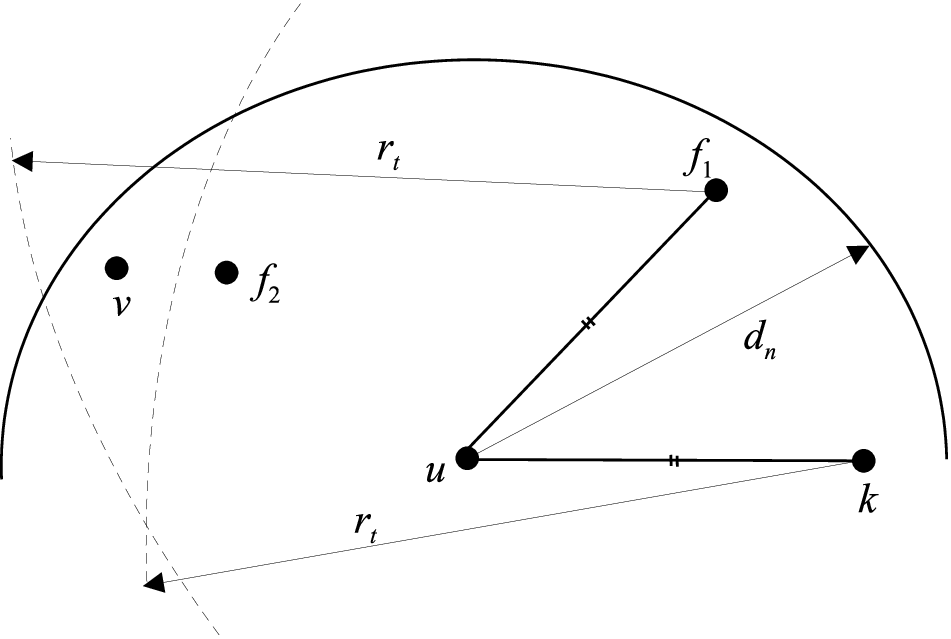,width=.7\textwidth,clip=}
\caption{Insufficient range for recognition of faulty node. Illustration
to the proof of Theorem~\ref{trmRange}.}\label{figRange}
\end{figure}

\Proof Consider the eventual neighborhood discovery and assume that
there is an algorithm $\cal A$ that solves the problem in the presence
of detectors on any layout without snares yet the transmission range
of the correct nodes $r_t$ is less than $2d_n$. Consider the layout
$L_1$ where the neighborhood of a correct node $u$ contains two nodes
$v$ and $f_1$ as well as a point $k$ with the following properties.
Refer to Figure~\ref{figRange} for illustration.  As usual, $v$ is
correct, $f_1$ is faulty and there is no node at point $k$. Even
though point $k$ is in the neighborhood of $u$, it is out of range of
$v$.  That is, $r_t < |vk|$.  Recall that this is possible since, by
assumption, $r_t < 2d_n$. Node $f_1$ is such that $|uf_1|=|uk|$ and
$r_t > |vf_1|$. The rest of the correct nodes in range of $u$ are
located such that, with the exception of $v$, $k$ forms a perfect
snare for $u$. That is, if $f_1$ sends a message from a fictitious
node $k$, the only node that generates conflict is $v$. Certainly,
with the presence of $v$, $k$ is not a snare so the assumptions of the
theorem apply.

Consider that $f_1$ indeed sends \emph{announcement} pretending to be
a fictitious node at $k$. Nodes $f_1$ and $k$ are equidistant from
$u$. Thus, if $f_1$ does not want $u$ to detect a conflict, $f_1$ has
to send the signal with the TSS of $T_r$. However, with such TSS, $v$
is in range of $f_1$ but out of range of $k$. This means that $v$
receives the announcement ostensibly coming from $k$ and detects a
conflict. The RSS at $v$ is $cT_r/|vf_1|^2$.  Since $\cal A$ is a
solution to the neighborhood discovery problem and $v$ is the only
node that is aware of the conflict, $v$ has to send \emph{conflict} to
$u$ which removes the fictitious node $k$ from the real universe of
$u$.

Consider a different layout $L_2$ (refer to Figure~\ref{figRange})
which is similar to $L_1$, only point $k$ is occupied by a correct
node and there is a faulty node $f_2$ near $v$. Specifically, the
distance $|vf_2|$ is such that there are no correct nodes within the
following range of $f_2$:
\[
\frac{|vf_2|}{|vf_1|}\sqrt{\frac{c}{R_{min}}}
\]
This ensures that when $f_2$ is going to imitate node $k$, none of
the nodes besides $v$ receive the messages from $f_2$. Note that $f_2$
and $k$ still do not form a snare because $v$ is aware of the
conflict. Note also, that such location of $f_2$ can always be found
if the faulty node can be placed arbitrarily close to $v$. 

Assume that if the node $k$ in $L_2$ sends a message, $f_2$ replicates
this message with TSS
\[
\frac{T_r|vf_2|^2}{|vf_1|^2}
\]
Observe that in this case all nodes, including $v$ and $u$, receive
exactly the same messages as in layout $L_1$. Since $\cal A$ is
deterministic, the nodes have to act exactly as in the previous
case. That is, $v$ has to issue a conflict with the message of node
$k$. However, after receiving this conflict, $k$ is separated from
$u$'s real universe. Recall that $k$ is correct in layout $L_2$. Note
that in this case $k$ is never going to be added to the output of $\cal
A$ at $u$. However, this violates the liveness property of the
neighborhood discovery problem since $k$ is a correct neighbor of $u$.
Thus, $\cal A$ is not a solution to this problem as we initially
assumed.  \QED

\section{The Sybil Attack Resilient Neighborhood Discovery
         Algorithm $\cal SAND$}\label{SecSAND}

Our description of the algorithm proceeds as follows. We first
motivate the need to frugally encode the universes to be passed to the
universes detectors. We then describe the operation of the
neighborhood detection algorithm itself. Then, we define the concrete
implementations of the abstract detectors specified in
Section~\ref{SecDetect}. These concrete detectors should operate with
our algorithm. On the basis of the algorithm and detector description
we state the theorem of algorithm correctness and detector optimality.

\noindent
\textbf{Encoding universes.} Observe that a na\"{i}ve solution
for representing universes by the algorithm results in an exponential
number of universes. Indeed, assume that node $u$ compiled a set of
nodes $U$ that do not conflict with two nodes $v$ and $w$. Suppose now
that $u$ records a conflict between the two nodes. They thus have to
be placed in separate universes: $U \cup \{v\}$ and $U\cup\{w\}$. Let
us consider another pair of conflicting nodes $x$ and $y$ that are
different from $v$ and $w$. Then, there are four possible universes:
$U \cup \{vx\}$, $U \cup \{vy\}$, $U \cup \{wx\}$, and $U \cup
\{wy\}$. Hence, if there are $N$ nodes in the neighborhood of $u$, the
potential number of conflicting pairs is $\lfloor N/2 \rfloor$ and the
number of universes is $2^{\lfloor N/2 \rfloor}$.

Therefore, our algorithm encodes the universes in the conflicts that
are passed to the detector. That is, the algorithm passes a set of
conflicts for the detector to generate the appropriate universe on its
own.

Recall also that in an asynchronous radio network the receiving node
can not distinguish one sender from another or decide if the two
messages were sent by the same node. This task has to be handled by
the detector.

\noindent
\textbf{Algorithm description.} We assume that the necessary
conditions for the existence of a solution to the neighborhood
discovery problem are satisfied: the layout does not contain a
(simple) snare and transmission range is at least twice as large as
the neighborhood distance $d_n$.

The $\cal SAND$ algorithm operates as follows. Every message
transmitted by the node contains its coordinates.  Each node sends
\emph{announce}. After receiving an \emph{announce}, a node replies
with a \emph{confirm} message. Each \emph{confirm} contains the
information of the announcement. If a node receives a message whose
coordinates do not match the received signal strength, the node
replies with a \emph{conflict} message. The \emph{conflict} also
contains the information of the message that generated the
conflict. Observe that \emph{confirm} can only be generated by
\emph{announce} while \emph{conflict} can be generated by an arbitrary
message.  Note that according to the locality assumption every node
ignores messages from the nodes outside of its neighborhood distance
$d_n$.

Each node $u$ builds a message dependency directed graph
\emph{DEP}. For each \emph{confirm}, $u$ finds a matching
\emph{announce}; for each \emph{conflict} --- a matching message that
caused the conflict. Note that this message dependence may not be
unique. For example a faulty node may send a message identical to a
message sent by a correct node.  Since a node cannot differentiate
senders in asynchronous radio networks, identical messages are merged
in \emph{DEP}. Note also, that a match may not be found because the
faulty node may send a spurious conflict message or the conflict
message is in reply to the faulty node message that $u$ does not
receive. Node $u$ removes the unmatched message. Also, $u$ removes the
cycles and sinks of \emph{DEP} that are not \emph{announce}. Observe
that \emph{DEP} may grow indefinitely as faulty nodes can continue to
send arbitrary messages.
 
Due to no-snare and transmission range assumptions, for every correct
process $u$ the following is guaranteed about \emph{DEP}:
\begin{itemize}
\item Eventually, $u$ receives an announcement from every correct node
in its neighborhood. An announcement from each correct node will be
confirmed by every correct node. There will be no messages from the
correct nodes that conflict with any other messages from the correct
nodes.
\item Eventually, every message from a fictitious node will be
followed up by at least one \emph{conflict} message sent by one of the
correct nodes from the neighborhood of $u$.
\end{itemize}

\noindent
\textbf{Concrete universe detectors.} We define the
\emph{concrete} detectors $c\cal SPU$, $c\cal WPU$ and $\diamond
c{\cal PU}$ as the detectors that accept the \emph{DEP} provided by
$\cal SAND$ as input and whose output satisfies the specification of
the corresponding abstract detectors described in
Section~\ref{SecDetect}. That is, for each correct node $u$, $c\cal
SPU$ only outputs complete and real universe, $c\cal WPU$ may output a
real universe that is not complete, while $\diamond c{\cal PU}$ may
provide arbitrary output for a fixed number of computation
states. However, all three detectors eventually output the complete
and real universe for $u$.  Observe that the detectors have to comply
with the specification even though \emph{DEP} may grow infinitely
large.

In $\cal SAND$, each process $u$ observes the output of the detector
and immediately outputs the universe presented by the detector without
further modification. By the construction of $\cal SAND$ proves the
following theorem.

\begin{theorem} \label{trmSAND} Considering layouts without simple 
snares and assuming that the transmission range is at least twice as
large as the neighborhood distance, the Sybil Attack Neighborhood
Detection Algorithm $\cal SAND$ provides a conflict-aware
deterministic solution to the Neighborhood Discovery Problem as
follows: $\cal SNDP$ if $c\cal SPU$ detector is used; $\cal WNDP$ if
$c\cal WPU$ is used; and $\diamond {\cal NDP}$ if $\diamond c{\cal PU}$
is used.
\end{theorem}

Similar to Chandra et al~\cite{CHT96} we can introduce the concept of
a weakest universe detector needed to solve a certain problem.  A
universe detector $\cal U$ is the \emph{weakest} detector required to
solve a problem $\cal P$ if the following two properties hold:
\begin{itemize}
\item there is an algorithm $\cal A$ that uses $\cal U$ to solve $\cal P$;
\item there is another algorithm $\cal B$ that uses the input of an
arbitrary solution $\cal S$ of $\cal P$ to implement $\cal U$.

\end{itemize}

\noindent
That is, $\cal B$ uses the output of $\cal S$ and provides the
computations expected of $\cal U$. The intuition is that if any
solution can be used to implement $\cal U$, then every solution needs
the strength of at least $\cal U$. Hence, the idea that $\cal U$ is
the weakest detector.

Observe that $\cal SAND$ provides the solutions using these detectors
to the respective problems.  Note also that the outputs of the
neighborhood discovery problems that we defined $\cal SNDP$, $\cal
WNDP$ and $\diamond\cal NDP$ can be used as the respective universe
detectors $\cal SPU$, $\cal WPU$ and $\diamond\cal PU$. For example,
if a process $u$ in $\cal SNDP$ outputs its neighborhood set, this
neighborhood set can be used to point to the real universe. Hence the
following proposition.

\begin{proposition}
Concrete universe detectors $c\cal SPU$, $c\cal WPU$ and $\diamond
c{\cal PU}$ are the weakest detectors required to solve $\cal SNDP$,
$\cal WNDP$ and $\diamond\cal NDP$ respectively.
\end{proposition}

\section{Detector Implementation and Future Research}\label{SecEnd}

\textbf{Detector implementation.} According to
Theorem~\ref{trmNotSolvable}, the universe detectors employed by our
solution to the neighborhood discovery problem are not themselves
implementable in asynchronous systems. The actual implementation of
the detectors can depend on the particular properties of the
application. Here are a few possible ways of constructing the
detectors. The nodes may be aware of the bounds on faulty nodes
speed. That is, the detectors will know the maximum number of
fictitious nodes they have to deal with.  The nodes may contain some
topological knowledge of the network. For example, the nodes may know
that the network is a grid. Alternatively, the nodes may have secure
communication with several trusted neighbors to ensure their presence
in the selected universe.

\noindent
\textbf{Future research.} We conclude the paper by outlining
several interesting areas of research that our study suggests. Even
though the concrete detectors we describe in the paper are minimal
from the application perspective, it is unclear if the input that
$\cal SAND$ provides is optimal. That is, is there any other
information that can be gathered in the asynchronous model that can help
the detector decide if a certain universe is real. We suspect that
$\cal SAND$ provides the maximum possible information but we would
like to rigorously prove it.

In this study, we assume completely reliable communication within a
certain radius of the transmitting node $R_{min}$. However, in
practice the propagation patterns of low-power wireless radios used in
sensor and other ad hoc networks are highly irregular. See for example
Zhou et al~\cite{ZHKS04}. The problem of adapting a more realistic
communication model is left open.

Another question is the true relationship between the universe and
fault detectors. Observe that unlike fault detectors, the universe
detectors require additional layout properties to enable the solution
to the neighborhood discovery. It would be interesting to research if
there is a complete analogue to fault detectors for this problem.

\bibliographystyle{plain} 
\bibliography{neighbors}

\end{document}